\documentclass[aps,prl,reprint,superscriptaddress]{revtex4-1}
\usepackage[utf8]{inputenc}

\usepackage{natbib,graphicx,amsmath,amssymb,bm,multirow,yfonts,color}
\usepackage[normalem]{ulem}
\newcommand{\be}{\begin{equation}}
\newcommand{\ee}{\end{equation}}
\newcommand{\bea}{\begin{eqnarray}}
\newcommand{\eea}{\end{eqnarray}}
\newcommand{\ba}{\begin{align}}
\newcommand{\ea}{\end{align}}

\renewcommand{\vec}[1]{\boldsymbol{#1}}

\begin{document}
\title{Muon creation in supernova matter facilitates neutrino-driven explosions}
\author{R.\ Bollig}
\affiliation{Max-Planck-Institut f\"ur Astrophysik, Karl-Schwarzschild-Str.~1, 85748 Garching, Germany}
\affiliation{Physik Department, Technische Universit\"at M\"unchen, James-Franck-Str.~1, 85748 Garching, Germany}
\author{H.-T.\ Janka}
\affiliation{Max-Planck-Institut f\"ur Astrophysik, Karl-Schwarzschild-Str.~1, 85748 Garching, Germany}
\author{A. Lohs}
\affiliation{GSI Helmholtzzentrum f\"ur Schwerionenforschung,
  Planckstra{\ss}e~1, 64291 Darmstadt, Germany}
\author{G. Mart{\'i}nez-Pinedo}
\affiliation{GSI Helmholtzzentrum f\"ur Schwerionenforschung,
  Planckstra{\ss}e~1, 64291 Darmstadt, Germany}
\affiliation{Institut f{\"u}r Kernphysik
  (Theoriezentrum), Technische Universit{\"a}t Darmstadt,
  Schlossgartenstra{\ss}e 2, 64289 Darmstadt, Germany}
\author{C.J. Horowitz}
\affiliation{Nuclear Theory Center and Department of Physics, Indiana
University, Bloomington, IN~47408}
\author{T. Melson}
\affiliation{Max-Planck-Institut f\"ur Astrophysik, Karl-Schwarzschild-Str.~1, 85748 Garching, Germany}


\begin{abstract}
Muons can be created in nascent neutron stars (NSs) due to the high 
electron chemical potentials and the high temperatures. Because of their
relatively lower abundance compared to electrons, their role has so far been
ignored in numerical simulations of stellar core collapse and NS formation.
However, the appearance of muons softens the NS equation of state, triggers
faster NS contraction and thus leads to higher luminosities and mean 
energies of the emitted neutrinos. This strengthens the postshock heating
by neutrinos and can facilitate explosions by the neutrino-driven mechanism.
\end{abstract}

\maketitle


{\em Introduction.---}First state-of-the-art three-dimensional (3D)
simulations have recently yielded successful supernova (SN)
explosions by the neutrino-driven mechanism
\cite{Takiwakietal2014,Melsonetal2015a,Melsonetal2015b,Lentzetal2015,Jankaetal2016,Robertsetal2016,Mueller2016,Muelleretal2017}.
But the explosions
turned out to be more delayed than in two-dimensional 
(axisymmetric; 2D) calculations and sensitive to neutrino
effects even on the 10--20\% level \cite{Melsonetal2015b}. 
Accurate physics in the neutrino and nuclear sectors is therefore
demanded to investigate the viability of the neutrino-driven
mechanism by self-consistent, first-principle 
neutrino-hydrodynamical simulations. 

While the presence of muons is well known to play a role in 
cold neutron stars (NSs; e.g.\ \cite{Steineretal2005,AlfordGood2010}),
it is traditionally ignored in SN matter
based on the argument that the high muon rest mass 
($m_{\mu}c^2 \approx 105.66$\,MeV) suppresses their 
formation. This reasoning, however, is not well justified
\cite{Lohsetal2014} because the electron chemical potential in 
newly formed NSs exceeds the muon rest mass, and
the peak temperatures rise above 30\,MeV after roughly 100\,ms 
after core bounce, when
the thermal distributions of photons and neutrinos reach well 
beyond 100\,MeV. These conditions enable the production of
significant numbers of muons and anti-muons ($\mu^-$, $\mu^+$)
via electromagnetic interactions such as $e^- + e^+ \longrightarrow
\mu^- + \mu^+$ and $\gamma + \gamma \longrightarrow
\mu^- + \mu^+$ ($\gamma$ denotes high-energy photons),
via weak reactions that couple the $e$-lepton and 
$\mu$-lepton sectors, and via $\beta$-processes
between nucleons and muon neutrinos
and antineutrinos ($\nu_\mu$, $\bar\nu_\mu$), which are created
in the SN core through thermal pair processes.

While the new-born NS loses
electron-lepton number by radiating a slight excess of electron 
neutrinos ($\nu_e$) compared to electron anti-neutrinos
($\bar\nu_e$), it also gradually builds up net muon-lepton
number (``muonizes'') by emitting more muon antineutrinos
than muon neutrinos. Electrons
and muons thus share the negative charge that compensates the
positive reservoir of protons (and of some $e^+$ and $\mu^+$).
Here we show that the rearrangements in the stellar medium and
the neutrino emission that are associated with the appearance
of muons have an important impact on the evolution of the 
proto-NS by accelerating its contraction. This facilitates
the development of SN explosions by the neutrino-driven
mechanism. Muons therefore must be included in self-consistent,
first-principle models of the SN phenomenon.


{\em Neutron star formation with muons.---}Assuming
neutrino-flavor oscillations do not play a role, conservation
equations for the lepton numbers (i.e., the numbers of the charged 
leptons plus their neutrinos minus those of the corresponding
anti-particles) for all three flavors hold individually.
During stellar core collapse neutrinos get trapped
and equilibrate at about one percent of the nuclear saturation density
($\rho_0\approx 2.7\times 10^{14}\,\mathrm{g\,cm}^{-3}$ or
baryon density $n_0\approx 0.16$\,fm$^{-3}$). 
Subsequently, they diffuse out of
the newly formed NS only over a time scale of several seconds.
The NS, which begins to form at core bounce, thus inherits a 
large concentration of electron-lepton number from the progenitor
core with an initial electron-flavor lepton
fraction of $\sim$$0.30$ electrons plus electron
neutrinos per baryon \cite{Mareketal2005}.
The diffusive loss of $\nu_e$ then
drives the evolution to the final neutron-rich state of a cold
NS with its small remaining content of protons.

In contrast, the trapped muon and tau-lepton numbers are zero
initially. The $\tau$ density remains extremely small at all
times because of the huge rest mass of the tauons
($m_{\tau}c^2 \approx 1777$\,MeV), which is much bigger than 
both the temperature and electron chemical potential in the
NS. Therefore the $\nu_\tau$ and $\bar\nu_\tau$ numbers are
initially equal and the chemical potentials $\mu_{\nu_\tau} = 
-\mu_{\bar\nu_\tau} = 0$ with high precision. However, 
since the cross section for neutral-current scattering with
nucleons, $\nu + N\longrightarrow\nu + N$ ($N = n,\,p$), 
is somewhat larger
for neutrinos than for anti-neutrinos due to
weak-magnetism corrections (of order $\epsilon/(m_Nc^2)$
with neutrino energy $\epsilon$ and nucleon mass $m_N$;
\cite{Horowitz2002}), $\bar\nu_\tau$ diffuse out faster and the 
proto-NS is expected to (transiently) develop a considerable 
tau-lepton number in the neutrino sector ($\mu_{\nu_\tau} > 0$) 
even though the formation of tauons is negligible
\cite{HorowitzLi1998}.

Different from tau neutrinos, but analogously
to $\nu_e$ and $\bar\nu_e$, $\nu_\mu$ and $\bar\nu_\mu$,
participate in $\beta$-reactions,
\begin{eqnarray}\label{eq:nuabsorption}
\nu_\ell+n\rightleftarrows p+\ell^-\,,\label{eq1}\\
\label{eq:antinuabsorption}
\bar\nu_\ell+p\rightleftarrows n+\ell^+\,,\label{eq2}
\end{eqnarray}
with their charged leptons, $\ell$ (standing for 
$e$ or $\mu$), when a significant population of thermally
excited $\mu^-$ and $\mu^+$ appears \cite{Lohsetal2014}. 
Beta equilibrium for both flavors implies the usual relation
\begin{equation}\label{eq:betaequilibrium}
\Delta \mu\equiv\mu_n-\mu_p=\mu_\ell-\mu_{\nu_\ell}
\end{equation}
between the chemical potentials (including particle
rest-mass energies) of neutrons, protons, charged leptons, 
and the corresponding neutrinos. Since the highly degenerate
Fermi sea of $e^-$ partially converts to $\mu^-$, and since
initially the trapped muon number is zero, an excess of 
$\mu^-$ over $\mu^+$ is compensated by an opposite excess
of $\bar\nu_\mu$ over $\nu_\mu$. Therefore the diffusive 
flux of $\bar\nu_\mu$ will dominate that of $\nu_\mu$,
leading to a gradual build-up of muon number.
The easier escape of $\bar\nu_\mu$ compared to $\nu_\mu$ is
aided by the lower neutral-current
scattering cross section for $\bar\nu_\mu$ mentioned above
and by the higher opacity for $\beta$-reactions of 
$\nu_\mu$ compared to $\bar\nu_\mu$ in analogy to the 
electron-flavor. The accumulation of net
muon number in the proto-NS, i.e.\ the process of muonization
that leads to an excess of $\mu^-$ over $\mu^+$ in the final NS,
is facilitated by the reactions of Eqs.~(\ref{eq1}) and
(\ref{eq2}). Also other interactions that couple the 
$e$-lepton and $\mu$-lepton sectors (Table~\ref{muonrates})
enhance the muonization rate and thus increase both the 
$\nu_\mu$ and $\bar\nu_\mu$ fluxes.

Muonization might play
a non-neglible role during all stages
of the SN post-bounce (p.b.) evolution and NS as well as
black-hole (BH) formation. 
In the following we discuss its effects on the initiation
of SN explosions by neutrino-energy deposition.

\begin{table}
\caption{Neutrino reactions with muons.\label{muonrates}}
\begin{ruledtabular}
\begin{tabular}{cc}
    $\quad \nu+\mu^{-}        \rightleftarrows \nu'+{\mu^{-}}'$ &
    $\nu+\mu^{+}              \rightleftarrows \nu'+{\mu^{+}}' \quad$ \\
    $\quad \nu_{\mu}+e^{-}    \rightleftarrows \nu_{e}+\mu^{-}$ &
    $\overline\nu_{\mu}+e^{+} \rightleftarrows \overline\nu_{e}+\mu^{+} \quad$ \\
    $\quad \nu_{\mu}+\overline\nu_{e}+e^{-} \rightleftarrows \mu^{-}$  &
    $\overline\nu_{\mu}+\nu_{e}+e^{+}       \rightleftarrows \mu^{+} \quad$ \\
    $\quad \overline\nu_{e}+e^{-} \rightleftarrows \overline\nu_{\mu}+\mu^{-}$ &
    $\nu_{e}+e^{+}           \rightleftarrows \nu_{\mu}+\mu^{+} \quad$ \\
    $\quad \nu_{\mu}+n       \rightleftarrows p+\mu^{-}$ &
    $\overline\nu_{\mu}+p    \rightleftarrows n+\mu^{+} \quad$
\end{tabular}
\end{ruledtabular}
\end{table}

\begin{figure}[t]
\includegraphics[width=\columnwidth]{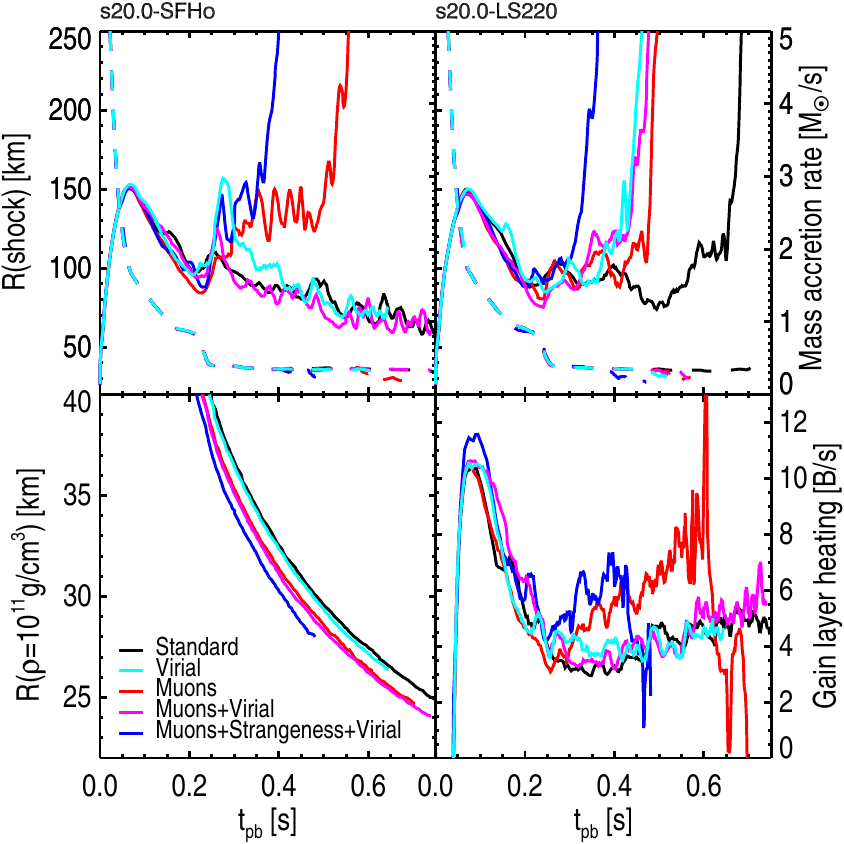}
\caption{{\em Upper row:} Angle-averaged shock radii (solid)
and mass-infall rates (at 400\,km; dashed) vs.\ post-bounce time for our sets
of models with SFHo ({\em left}) and LS220 EoS ({\em right}).
{\em Lower row:} Time evolution of NS radii (measured at an average density
of $10^{11}$\,g\,cm$^{-3}$; {\em left}) and net heating rate integrated over
the gain layer (in $1\,\mathrm{B\,s}^{-1} = 
10^{51}$\,erg\,s$^{-1}$; {\em right}) for models with SFHo EoS.}
\label{fig:shockradii}
\end{figure}

\begin{figure}[t]
\includegraphics[width=\columnwidth]{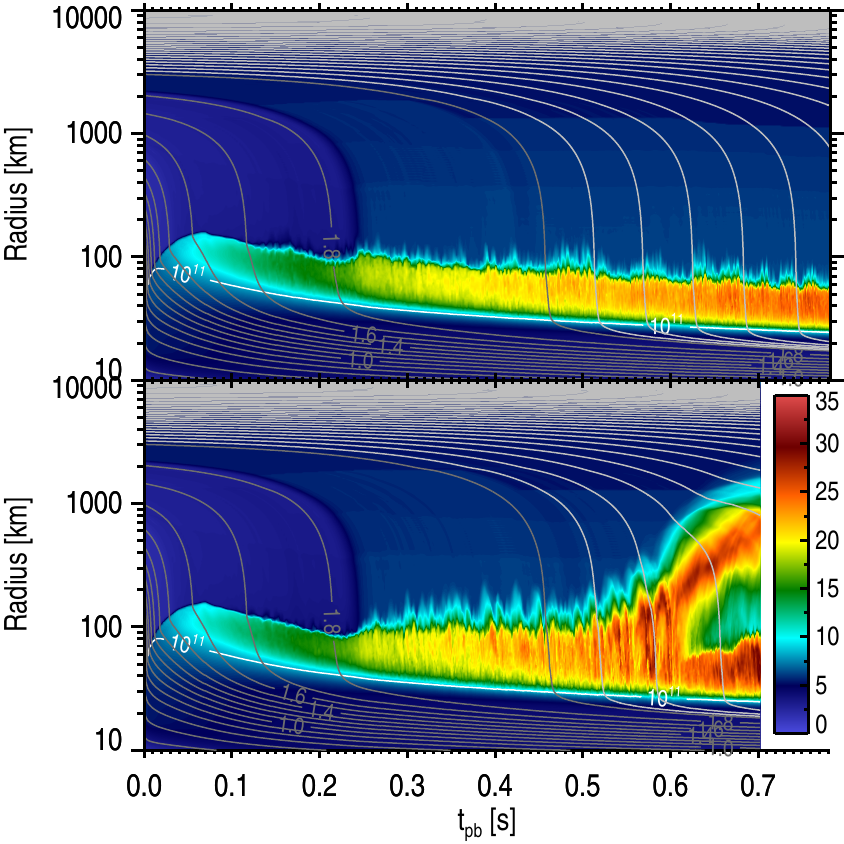}
\caption{Evolution of angle-averaged specific entropy (color; in $k_\mathrm{B}$
per nucleon) vs.\ post-bounce time for model s20.0-SFHo with our standard
physics ({\em top}) and with muons ({\em bottom}). The outer boundary of
the light blue-green-yellow region follows the average radius of the SN shock.
The gray lines mark
``mass shells'' (radii of constant enclosed baryonic mass), the white line
corresponds to an average density of $10^{11}$\,g\,cm$^{-3}$.}
\label{fig:massshells}
\end{figure}


{\em Numerical modeling.---}Our SN simulations were performed
with the PROMETHEUS-VERTEX neutrino-hydrodynamics 
code \cite{RamppJanka2002,Burasetal2006}
with an approximate treatment of general
relativistic gravity by the effective gravitational potential
of Case~A of \cite{Mareketal2006}.
The PROMETHEUS hydrodynamics module solves the
equations of non-relativistic hydrodynamics (continuity 
equations for mass, momentum, energy, lepton number, and
nuclear composition) with an explicit, directionally-split,
higher-order Godunov scheme \cite{Fryxelletal1989}.
The transport module VERTEX integrates the energy-dependent
evolution equations of energy and momentum 
for all six neutrino species ($\nu_e$,
$\bar\nu_e$, $\nu_\mu$, $\bar\nu_\mu$, $\nu_\tau$, $\bar\nu_\tau$)
in the comoving frame of the stellar fluid to order $v/c$ ($v$ is
the fluid velocity, $c$ the speed of light), including
corrections due to general relativistic redshift and time dilation.
The closure is provided by an Eddington factor based on the solution
of a model-Boltzmann equation, iterated for convergence with the 
set of two-moment equations \cite{RamppJanka2002}. 
Neutrino transport in multi-dimensional simulations employs the
ray-by-ray plus approximation \cite{Burasetal2006}.

We upgraded the PROMETHEUS-VERTEX code for 
including all effects of $\mu^-$ and $\mu^+$ in the hydrodynamics
and equation of state (EoS) of the stellar plasma, the effective
relativistic gravity potential, and in the neutrino transport.
This implies the solution of conservation equations for
electron and muon lepton number: 
\begin{equation}
\frac{\partial(\rho Y_\ell)}{\partial t} + {\vec\nabla}(\rho Y_\ell\,\vec v) 
= Q_\ell
\label{eq:deltaY}
\end{equation}
(here relativistic corrections are omitted for simplicity).
$Y_\ell = Y_{\ell^-} - Y_{\ell^+}$ is the net number of charged
leptons per nucleon, $\rho$ the baryon-mass density
and $Q_\ell$ the source rate that is associated with
all processes emitting and absorbing $\nu_\ell$ and $\bar\nu_\ell$.
The EoS depends on $Y_e$ and $Y_\mu$, i.e., 
$P = P(\rho,T,Y_e,Y_\mu,\{Y_k\}_{k = 1,...,N_\mathrm{nuc}})$ and
$\omega = \omega(\rho,T,Y_e,Y_\mu,\{Y_k\}_{k = 1,...,N_\mathrm{nuc}})$
for pressure $P$ and specific energy density $\omega$ ($T$ is the
medium temperature, $N_\mathrm{nuc}$ the number of nuclear species). 
Analogously 
to $e^-$ and $e^+$, $\mu^-$ and $\mu^+$ provide an additive 
contribution to $P$ and $\omega$ and are treated as ideal Fermi
gases of arbitrary degeneracy and arbitrary degree of relativity.
In nuclear statistical equilibrium (NSE)
the mass fractions of nuclei and nucleons, $Y_k$, are determined
by the Saha equations and hence $Y_k = Y_k(\rho,T,Y_e,Y_\mu)$ holds;
otherwise they follow from evolution equations similar 
to Eq.~(\ref{eq:deltaY}) with $Q_\ell$ being replaced by source
terms for nuclear reaction rates. With $\rho$, $\omega$, $Y_e$ and 
$Y_\mu$ given as solutions of the hydrodynamics and $Y_k$
($k = 1,...,N_\mathrm{nuc}$) being determined either 
by NSE or Eq.~(\ref{eq:deltaY}), $T$ and the chemical potentials 
$\mu_e$, $\mu_\mu$, $\mu_n$, $\mu_p$, and $\mu_k$ for all $k$
can be determined under the constraint of charge neutrality,
$\sum_k Z_kY_k = Y_e + Y_\mu$, with $Z_k$ being the nuclear charge
number of species $k$.

Accounting for the presence of muons and the differences of
the $\nu$ and $\bar\nu$ scattering cross sections with nucleons
due to nucleon-recoil and weak-magnetism \cite{Horowitz2002},
we generalized the neutrino-transport module VERTEX to an 
energy-dependent six-species treatment, tracking $\nu_e$, 
$\bar\nu_e$, $\nu_\mu$, $\bar\nu_\mu$, $\nu_\tau$, and $\bar\nu_\tau$
individually. Besides our ``standard'' set of neutrino reaction
rates listed in Table~1 of \cite{Janka2012}, we also implemented
all relevant neutrino interactions with $\mu^-$ and $\mu^+$ as
listed in Table~\ref{muonrates}. The detailed kinematics (energy
and momentum exchange between reaction partners) were fully taken 
into account, describing charged leptons as arbitrarily
relativistic and arbitrarily degenerate fermions and nucleons
as non-relativistic fermions. Neutral and charged-current 
interactions between neutrinos and nucleons were handled by
the formalism of \cite{BurrowsSawyer1998,BurrowsSawyer1999},
which includes the effects of nucleon correlations by a
random-phase approximation (RPA). We generalized the treatment 
to also include corrections due to neutron and proton
mean-field potentials in the $\beta$-processes
\cite{Reddyetal1998,MartinezPinedoetal2012,Robertsetal2012} and due
to the large rest masses of $\mu^-$ and $\mu^+$. 
Weak-magnetism
corrections according to \cite{Horowitz2002} are used in all
neutral and charged-current neutrino-nucleon interactions 
(cf.~\cite{Burasetal2006}) except in charged-current reactions
of $\nu_\mu$ and $\bar\nu_\mu$ with nucleons (because lepton-mass
dependence was neglected in \cite{Horowitz2002}).
Neutral and charged-current reactions 
of neutrinos with nucleons bound in light nuclei ($^2$H, $^3$H, 
$^3$He) were approximated by using the neutrino-nucleon interactions 
of \cite{Bruenn1985}, which slightly overestimates (mainly at 
low energies) the collective opacity of these reactions compared to
the detailed description in \cite{Fischeretal2016}.
When specified, we included in neutrino-nucleon scatterings 
virial corrections for the axial response of nuclear matter at
low densities \cite{Horowitzetal2017,LinHorowitz2017} and/or applied  
a strangeness-dependent contribution to the axial-vector coupling
coefficient \cite{Horowitz2002} with a value of 
$g_\mathrm{A}^\mathrm{s} = -0.1$, consistent with experimental
constraints \cite{Hobbsetal2016}. The virial corrections were
implemented via an effective interaction in the RPA that 
was stronger at low densities. This yielded results similar to 
those in \cite{Horowitzetal2017}.

Our SN simulations were performed in 2D for a non-rotating 
20\,$M_\odot$ progenitor model \cite{WoosleyHeger2007} with the
Lattimer-Swesty EoS (LS220) with nuclear incompressibility
$K = 220$\,MeV \cite{LattimerSwesty1991} and the SFHo EoS
\cite{Hempeletal2012,Steineretal2013} (models s20.0-LS220
and s20.0-SFHo, respectively). After bounce, at densities below 
$10^{11}$\,g\,cm$^{-3}$, we employed a 23-species NSE solver
at $T > 0.5$\,MeV for infalling and $T > 0.34$\,MeV for 
expanding, high-entropy matter, and nuclear ``flashing'' 
\cite{RamppJanka2002} at lower temperatures. For the polar 
coordinate grid we used a time-dependent
number of 400--650 radial zones and 160 lateral zones with a
refinement to 320 lateral zones outside of the gain radius
(i.e., the radius exterior to which neutrino heating dominates),
and for the neutrino transport 15 geometrically distributed
energy bins with $\epsilon_\mathrm{max} = 380$\,MeV.

\begin{figure*}[h!t]
\includegraphics[width=.87\textwidth]{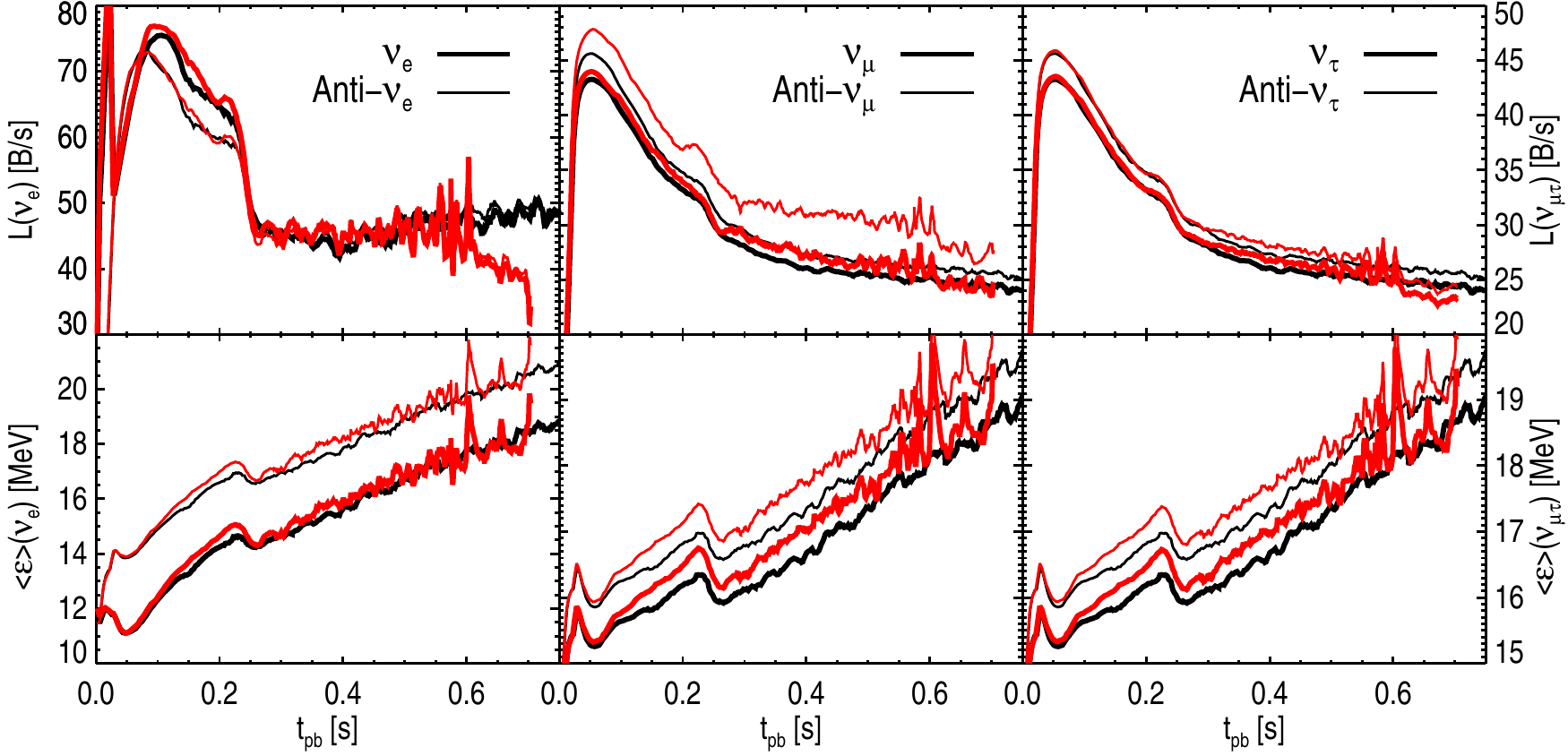}
\caption{Neutrino luminosities 
({\em upper row}) and radiated mean neutrino energies (defined as
ratio of neutrino energy density to number density; {\em lower row}) 
vs.\ post-bounce
time, evaluated in the laboratory frame at the average gain radius
for the standard model with SFHo EoS (black) and the simulation with
muons (red).}
\label{fig:nusignal}
\end{figure*}

\begin{figure}[t]
\includegraphics[width=\columnwidth]{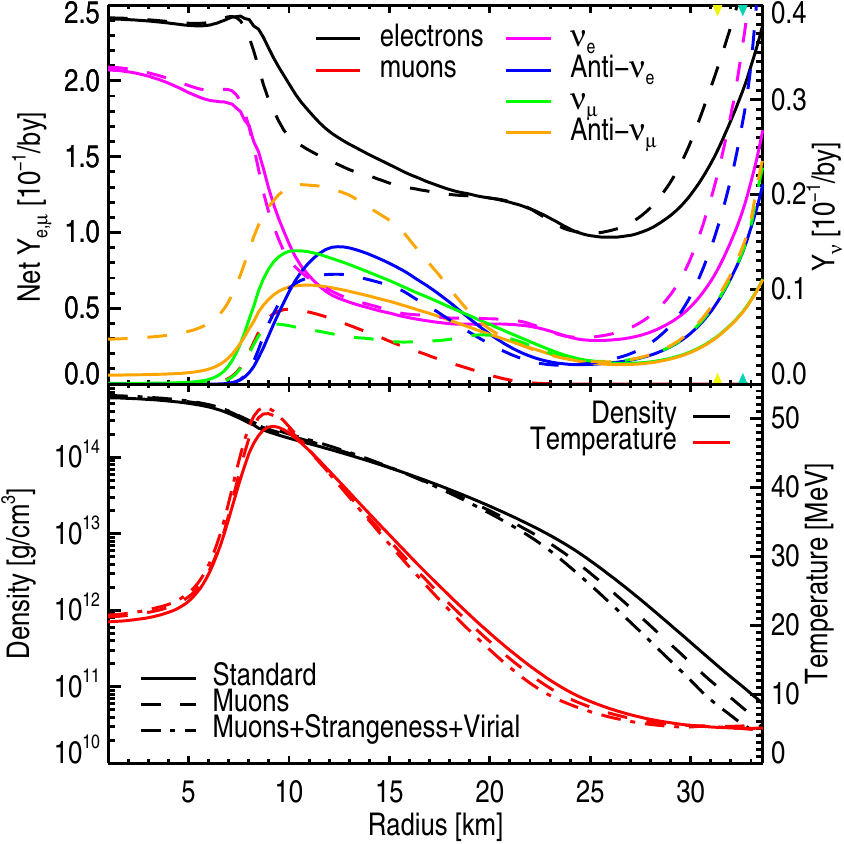}
\caption{{\em Top:} Radial profiles at 0.4\,s after bounce of the 
net numbers of charged leptons (left-hand scale) and neutrinos 
(right-hand scale) per baryon for the standard
model with SFHo EoS (solid) and the simulation with muons (dashed). 
{\em Bottom:} Radial profiles of density (black) and temperature
(red) for three cases with SFHo EoS.}
\label{fig:profiles}
\end{figure}


{\em Results.---}In addition to
conducting simulations for the two employed nuclear 
EoSs with our standard set of neutrino processes
(Table~1 in \cite{Janka2012}), we also investigated cases where 
we included (a) the virial corrections in $\nu$-$N$-scattering, 
(b) all muon effects, (c) both muon and virial effects, and (d) muons,
virial effects, and a strangeness correction in $\nu$-$N$-scattering.
Figure~\ref{fig:shockradii} displays the time evolution of the average 
shock radii for the models with SFHo (top left) and LS220 EoS (top
right). It is obvious that muon formation enables an explosion for
the SFHo model, which does not explode with standard neutrino 
physics, and it allows for an earlier onset of the explosion 
with the LS220 EoS. 

Figure~\ref{fig:massshells} compares the evolution
of angle-averaged radial profiles of the entropy per baryon 
(superimposed in color on mass-shell trajectories) for two 
SFHo-models.
After the arrival of the interface between silicon-shell and 
oxygen-rich Si-layer at the shock at $\sim$240\,ms p.b.,
the shock radius in the model with muons is considerably larger
than in the standard case, leading to an explosion, despite
the inverse order of the shock radii at earlier times 
(Fig.~\ref{fig:shockradii}).
The lower panels of Fig.~\ref{fig:shockradii} provide an
explanation: with muons the proto-NS contracts notably faster
(left). The creation of $\mu^-$ and $\mu^+$ effectively softens 
the EoS by conversion of thermal and degeneracy energy of $e^-$
into rest-mass energy of muons. In addition, it significantly 
raises the emission of $\bar\nu_\mu$ and, 
to a lesser extent, also
of $\nu_\mu$ (Fig.~\ref{fig:nusignal}, middle panels). The
accelerated shrinking of the NS leads
to higher temperatures at given densities and correspondingly
increased luminosities and mean energies of the emitted 
electron- and tau-flavor neutrinos, which
are shown in Fig.~\ref{fig:nusignal} (left and right panels)
at the gain radius, where $\nu_e$ and $\bar\nu_e$ differences
are relevant for the neutrino
heating. As a consequence, the neutrino-heating rate, per baryon
as well as integrated over the gain layer (i.e.\ the
region between gain radius and shock), becomes sizeably greater
in the model with muons at $t\gtrsim 240$\,ms 
(Fig.~\ref{fig:shockradii}, bottom right). Muons therefore
have a similar overall effect as the strangeness-dependent
reduction of neutrino-nucleon scattering discussed in
\cite{Melsonetal2015b}.

Figure~\ref{fig:profiles} documents
the appearance of significant charged-muon number
(up to $Y_\mu\sim 0.05$) (at the expense of $e^-$) correlated
with a temperature maximum in the NS 
between $\sim$7\,km ($\sim$$4\times 10^{14}$\,g\,cm$^{-3}$) and
$\sim$21\,km ($\sim$$2\times 10^{13}$\,g\,cm$^{-3}$). While in the
model without muons $\nu_\mu$ are more abundant than $\bar\nu_\mu$,
equivalent to the situation for $\nu_\tau$ and $\bar\nu_\tau$
discussed above, the situation is reversed
when muons are included: $Y_{\nu_\mu}$ drops in its peak to about 
half of the abundance in the standard case, whereas the number
of $\bar\nu_\mu$ more than doubles ($Y_{\bar\nu_\mu}^\mathrm{max} 
\gtrsim 0.02$).

Including also strangeness corrections in $\nu$-$N$ scattering
leads to even faster explosions (Fig.~\ref{fig:shockradii}, upper
panels), because muon and strangeness effects drive the system in
the same direction, namely a faster contraction of the NS 
(Fig.~\ref{fig:shockradii}, bottom left). The situation for
virial effects is ambiguous. While the LS220 model with virial
corrections explodes faster than the standard case and evolves
similar to the simulation with muons, virial effects in addition
to muons make little difference (Fig.~\ref{fig:shockradii}, top
right). In contrast, an SFHo model including virial 
corrections and strangeness $g_\mathrm{A}^\mathrm{s} = -0.1$
(not shown) explodes only later than 600\,ms due to the
strangeness effects, whereas the
SFHo models with virial response fail to 
explode with and without muons (Fig.~\ref{fig:shockradii}, top left).
For relevant temperatures ($T \approx 5$--10\,MeV)
virial effects lead to a reduction of the 
$\nu$-$N$-scattering opacity compared to RPA results only at 
densities below $\sim$$(0.01\,...\,0.03)\rho_0$. This is so low
that there is a visible (1--2\%) increase of the heavy-lepton 
neutino emission but hardly any correspondingly accelerated
contraction of the NS
radius (Fig.~\ref{fig:shockradii}, bottom left). Virial 
effects are therefore subtle, because they can 
enhance energy extraction in the $\nu_\mu$ and
$\nu_\tau$ sector without explosion-favoring consequences for
emission and heating by $\nu_e$ and $\bar\nu_e$.

{\em Conclusions.---}We
have demonstrated by 2D simulations that the appearance of
muons in the hot medium causes enhanced neutrino emission and
faster contraction of the proto-NS with supportive effects on
the neutrino-energy deposition behind the stalled shock and the
onset of neutrino-driven explosions. The ongoing muonization of
the new-born NS may also lead to stronger heating of matter
that is still accreted and re-ejected after the onset of the 
explosion (see \cite{Muelleretal2017} and references therein)
and could therefore raise the explosion energy. Muonization
mainly affects more massive and thus hotter NSs and should have
less impact on SN explosions with less massive
NSs. Final conclusions about their detailed role in
the explosion will require 3D simulations. Since muon
formation effectively softens the NS EoS at high densities, it
also has important implications for the collapse of hot NSs
to BHs \cite{Summaetal2017}. Therefore muons cannot be ignored 
in detailed models of the SN explosion mechanism and NS formation.
For a rigorously self-consistent description,
this requires ---and we have implemented--- a full six-species
treatment of neutrino transport, which couples the production
of electron- and muon-flavor neutrinos.
Since all six neutrino species differ
in their spectra, corresponding transport results may 
offer interesting new aspects for neutrino oscillations.
Muons may also have to be included in simulations
of NS-NS mergers, because the compactness of the merger
remnant and its time scale for a possible collapse to a BH
is sensitive to muon formation in the hot nuclear medium.

\begin{acknowledgments}
We acknowledge hospitality by ECT$^\ast$ in Trento during the
MICRA 2013 workshop, where discussions helped start this work.
At Garching, this work was supported by the European Research
Council through grant ERC~AdG~341157-COCO2CASA and by the
Deutsche Forschungsgemeinschaft through grants SFB~1258 
(``Neutrinos, Dark Matter, Messengers'') and EXC~153
(``Excellence Cluster Universe''). GMP acknowledges partial
support by the Deutsche Forschungsgemeinschaft through grant
SFB~1245 (``Nuclei: From Fundamental Interactions to Structure 
and Stars''). CJH was supported in part by US DOE Grants 
DE-FG02-87ER40365 and DE-SC0008808.
The numerical simulations were carried out at the
Max Planck Computing and Data Facility (MPCDF).
\end{acknowledgments}

\bibliographystyle{h-physrev5}
\bibliography{references}

\end{document}